# Preparation of the silver boride: preliminary results

M. Sinder and J. Pelleg

*Department of Materials Engineering, Ben - Gurion University of the Negev, Beer Sheva, 84105, Israel*

## Abstract

Bulk and thin film specimens of $AgB_2$ were prepared. The bulk pellets were obtained from the powders of the constituents and the films were produced by cosputtering and sequential sputtering. The specimens were annealed and subjected for X-ray and Auger analysis. Preliminary results seems to indicate that $AgB_2$ is an instable phase.

## Introduction

The discovery of the new superconductor $MgB_2$ with critical temperature $T_c$ = 39 K [1] stimulated great activity in search for materials with higher $T_c$. Some theoretical works predicted the possibility of obtaining such supercomputing materials [2]. However to date, the experimental attempts to find materials with higher $T_c$ than 39 K failed [3, 4].

Theoretical investigations of diborides of metals having the $AlB_2$ type structure predicted the existence of superconductivity in $AgB_2$ with a $T_c$ higher than 39K [5-8]. $AgB_2$ seemed to be a compound, which is difficult to synthesize and only one successful experimental work is known [9,10]. The results of this work appeared in a short note claiming successful preparation of $AgB_2$. Its structure was determined as being that of AlB$_2$ and the lattice parameters of this compound were evaluated as a = 0.300 nm and c = 0.324 nm.

In this communication we report our preliminary results of efforts to prepare the silver boride phase.

**Experiment**

We have tried three experimental approaches to produce silver boride: a) powder approach, b) cosputtering of Ag and B and c) sequential sputtering of Ag and B.

**a) Powder approach:**

Ag and B powders were mixed in a ratio Ag : B = 1 : 2 and pressed into pellets. Each pellet was sealed in a quartz ampoule under vacuum and backfilled with a small amount of Ar. Annealing in a box furnace was performed at 1173 K for the times of 2, 5, 9 and 24 h. Ampoules were quenched in water to room temperature. X-ray diffraction (XRD) analysis of these samples was performed by a Rigaku diffractometer.

**b) Cosputtering of Ag and B:**

Magnetron cosputtering from Ag and B targets on Si (100) substrate in Ar atmosphere was used for the deposition. The base pressure prior to film deposition was ~ 10$^{-5}$ Pa. RF power was applied on both targets. The film was annealed in ampoules of quartz at 933 K and then analyzed by XRD and Auger techniques.

**c) Sequential sputtering of Ag and B:**

B and Ag were successively deposited by magnetron sputtering from the respective targets on Si (100) substrate respectively. The sputtering conditions were the same as those for the cosputtering. The resulting film was annealed at 933 K in quartz ampoules and then analyzed by XRD technique.

**Results**

a) **Powder approach**

The XRD spectrum of the specimen annealed at 1173 K for 9 hours is shown in Fig. 1. Two $AgB_2$ peaks of the (001) and (002) reflections were observed in addition to those which are related to the Ag phase. It was observed that the intensity of peaks increased with increasing annealing time, but a prolonged annealing of 24 h resulted in the disappearance of the $AgB_2$ peaks. We do not have a good explanation for this observation except the possibility that the phase is not stable at this high temperature for extended times. It should be noted that the specimen annealed for 9 h was retested after a year and the results were the same, i.e. the $AgB_2$ peaks remained. We interpret the results as indicating stability of the $AgB_2$ phase at room temperature, but instable when exposed to high temperature for long time. From these peaks the c parameter was evaluated as c = 0.318 nm. This value is very close to the one report in reference [9]

b) **Cosputtering of Ag and B**:

Fig. 2 is an illustration of an XRD spectrum of a specimen obtained by cosputtering and annealed at 933 K for 30 min. Only the (100) $AgB_2$ reflection was observed thus confirming the existence of the $AgB_2$ phase in the cosputtered film also. An Auger spectrum of a specimen cosputtered under the same conditions as the specimen illustrated in Fig. 2 but annealed at 933 K for 4 min is shown in Fig. 3. Note that the Auger spectrum does not show the ratio between Ag and B as could be expected by the composition of the $AgB_2$ phase. In one of the thin film specimens the $AgB_2$ peak disappeared after several days at room temperature.

**c) Sequential sputtering of Ag and B:**

Fig. 4 is an XRD spectrum of a film obtained by sequential sputtering and annealing at 933 K for 1 h. Again only one $AgB_2$ reflection from the (001) plane was observed. This specimen was retested after several days and the (001) peak could not be detected.

A common observation of some of the films obtained by sputtering was that the $AgB_2$ peaks disappear after a few days. In bulk specimens the $AgB_2$ peaks disappeared only when exposed to high temperature for long times. We do not have yet a satisfactory explanation of these observations. If for the sake of this note the possibility of experimental artifacts are excluded, a logical speculation could be related to the relative stability of $AgB_2$. Thus, bulk $AgB_2$ specimens require high temperature exposure for long times to be decomposed, whereas in thin films this same phenomenon can occur already at room temperature exposure for a few days. This puzzling observation of instability requires a thorough study of the subject. Our current attempts are devoted to understand this instability in $AgB_2$ specimens. In addition the superconductive behavior of the stable bulk specimens is currently investigated.

**Summary**

Silver boride is a phase, which can be synthesized, but its stability as a function of time at high temperature or low temperature is not yet clear. $AgB_2$ bulk specimens seem to be an instable phase when exposed to high temperature for long times. The instability of thin films is observed at room temperature when exposed for a few days.

The time and temperature dependent instability of AgB$_2$ specimens is being studied currently.

**Acknowledgement**

# List of Figures

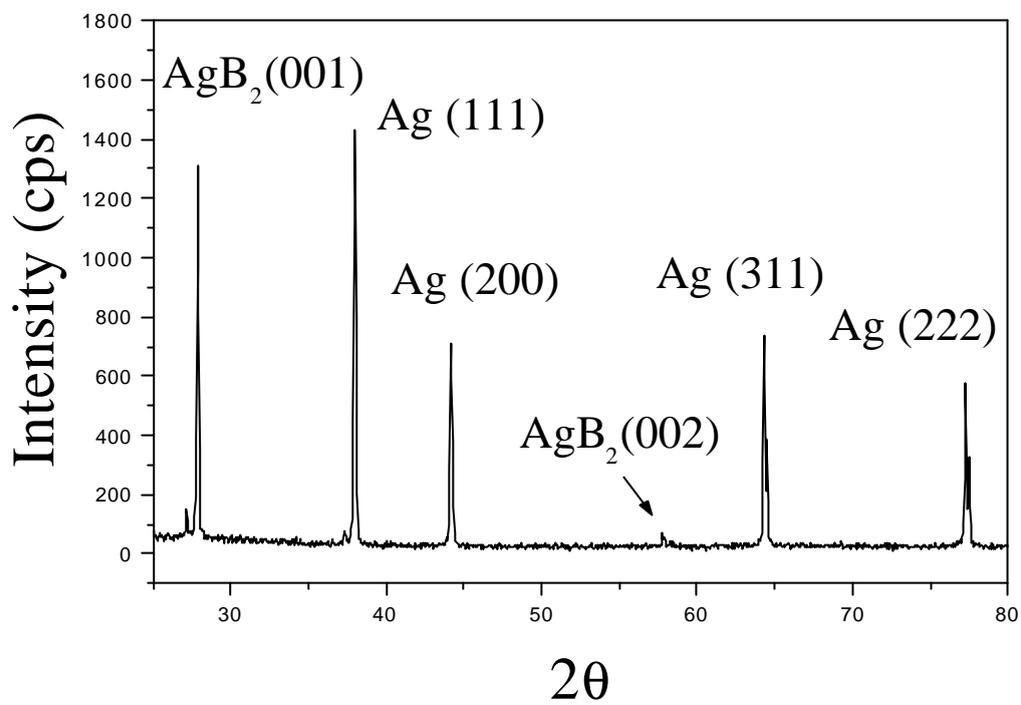

Fig. 1.

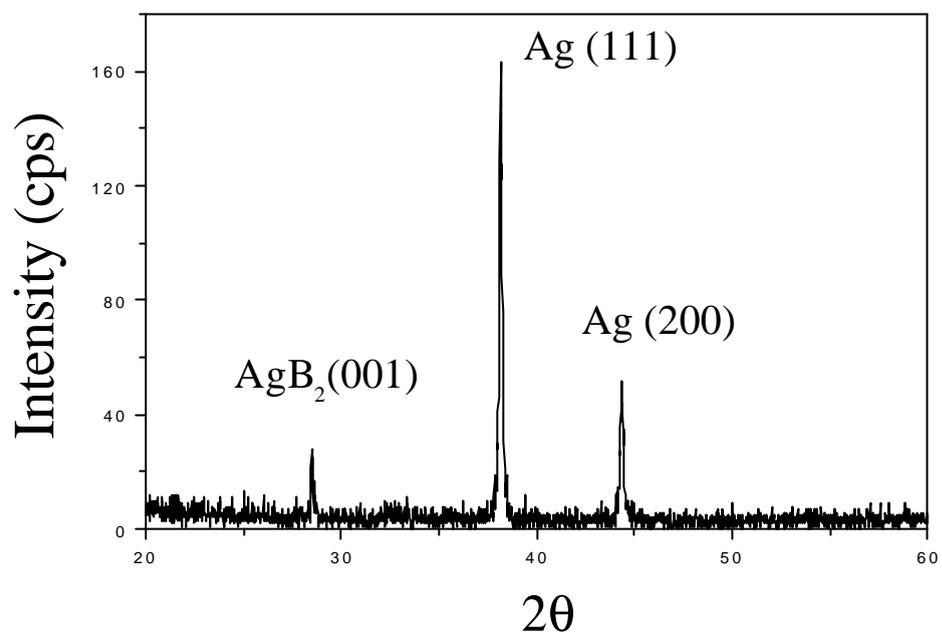

Fig. 2.

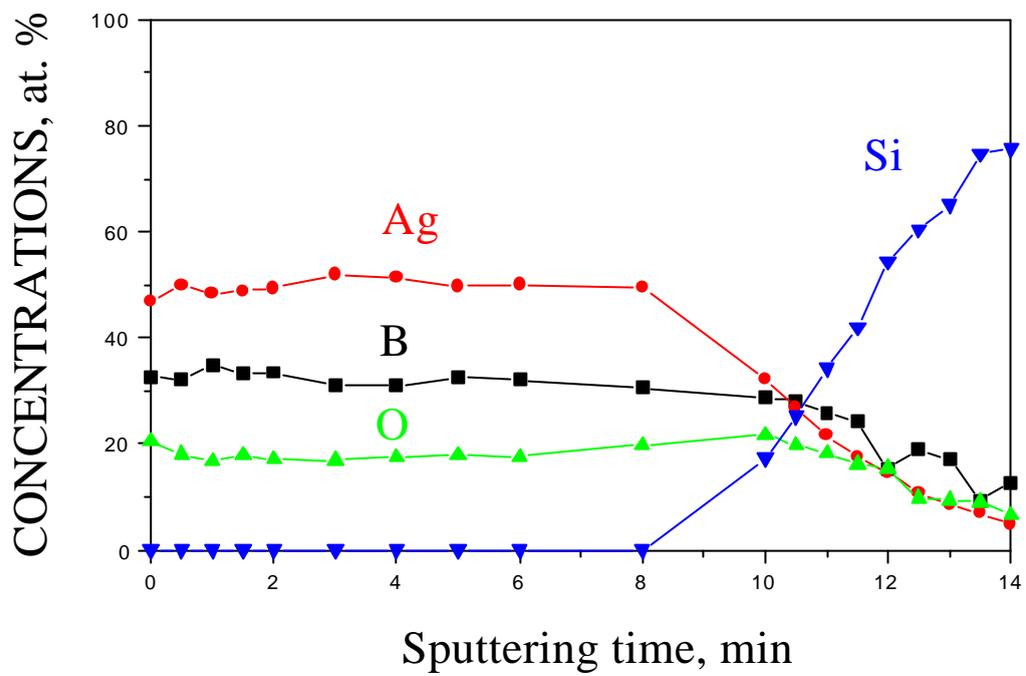

Fig. 3.

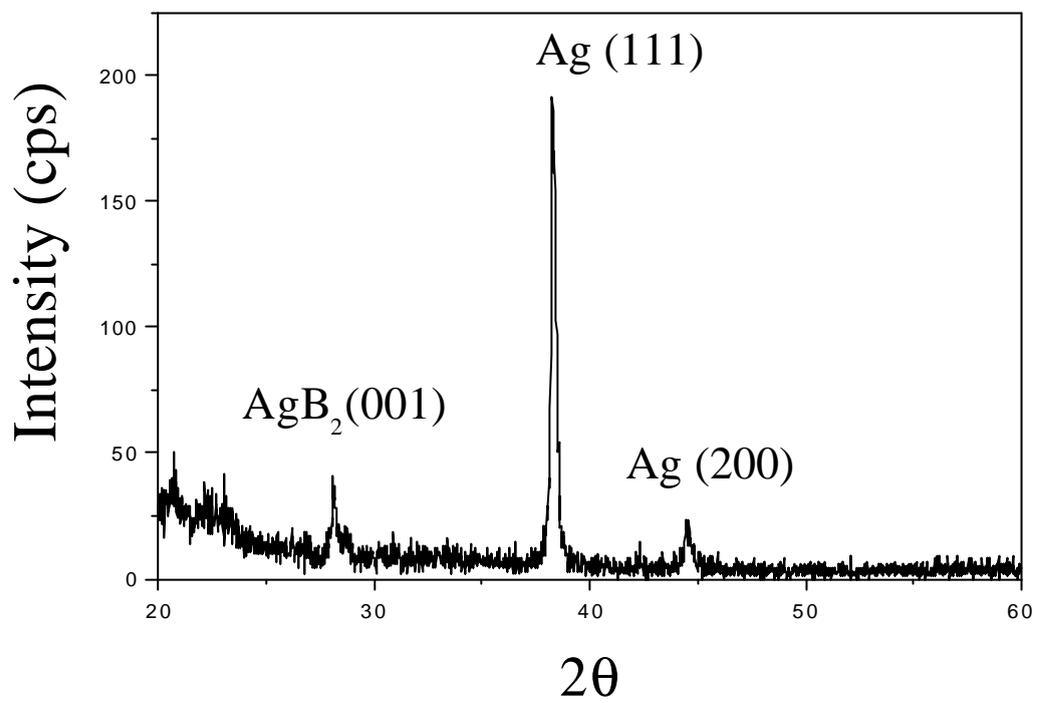

Fig. 4.